\documentclass[10pt,english,two column]{IEEEtran}
\renewcommand{\baselinestretch}{0.95}
\usepackage[T1]{fontenc}
\usepackage[latin9]{inputenc}
\usepackage{float}
\usepackage{amsmath}
\usepackage{amsthm}
\usepackage{amssymb}
\usepackage{graphicx}
\usepackage{multirow}
\usepackage{xcolor}
\usepackage{subcaption}
\usepackage{multicol}
\makeatletter
\floatstyle{ruled}
\newfloat{algorithm}{tbp}{loa}
\providecommand{\algorithmname}{Algorithm}
\floatname{algorithm}{\protect\algorithmname}

\theoremstyle{plain}

\theoremstyle{plain}

\ifCLASSINFOpdf% \usepackage[pdftex]{graphicx}
\else% or other class option (dvipsone, dvipdf, if not using dvips). graphicx
\fi% graphicx was written by David Carlisle and Sebastian Rahtz. It is
\usepackage{babel}

\makeatother

\usepackage{babel}
\providecommand{\propositionname}{Proposition}
\providecommand{\theoremname}{Theorem}

\begin{document}

\title{\huge{Deep Learning Methods for Universal MISO Beamforming}}

\author{Junbeom Kim, Hoon Lee, Seung-Eun Hong and Seok-Hwan Park \thanks{This work was supported by Basic Science Research Program through the National Research Foundation of Korea (NRF) grants funded by the Ministry of Education {[}NRF-2019R1A6A1A09031717{]}. This work also supported by the Pukyong National University Research Fund in 2019 (CD20191034) and by Institute for Information \& communications Technology Promotion (IITP) grant funded by the Korea government (MSIT) (No.2018-0-01410, Development of Radio Transmission Technologies for High Capacity and Low Cost in Ultra Dense Networks).

J. Kim and S.-H. Park are with the Division of Electronic Engineering, Jeonbuk
National University, Jeonju, Korea (email: \{junbeom, seokhwan\}@jbnu.ac.kr).

H. Lee is with the Department of Information and Communications Engineering, Pukyong National University, Busan, Korea (email: hlee@pknu.ac.kr).

S.-E. Hong is with Electronics and Telecommunications Research Institute (ETRI), Daejeon, Korea (email: iptvguru@etri.re.kr).}}
\maketitle

\begin{abstract}
This letter studies deep learning (DL) approaches to optimize beamforming vectors in downlink multi-user multi-antenna systems that can be universally applied to arbitrarily given transmit power limitation at a base station. We exploit the sum power budget as side information so that deep neural networks (DNNs) can effectively learn the impact of the power constraint in the beamforming optimization. Consequently, a single training process is sufficient for the proposed universal DL approach, whereas conventional methods need to train multiple DNNs for all possible power budget levels. Numerical results demonstrate the effectiveness of the proposed DL methods over existing schemes.
\end{abstract}


\theoremstyle{theorem}
\newtheorem{theorem}{Theorem}
\theoremstyle{proposition}
\newtheorem{proposition}{Proposition}
\theoremstyle{lemma}
\newtheorem{lemma}{Lemma}
\theoremstyle{corollary}
\newtheorem{corollary}{Corollary}
\theoremstyle{definition}
\newtheorem{definition}{Definition}
\theoremstyle{remark}
\newtheorem{remark}{Remark}

\vspace{-3mm}
\section{Introduction}
% \vspace{-1mm}

In the last decades, there have been intensive studies on linear beamforming techniques for multi-antenna multi-user systems. Advances in the non-convex optimization theory have brought systematic optimization strategies for multi-antenna signal processing \cite{Bjornson-et-al:SPM14}. Existing handcraft beamforming optimization algorithms generally require significant computing power. In particular, weighted minimum mean square error (WMMSE) algorithm \cite{Christensen-et-al:TWC08} obtains a locally optimal beamforming solution for broadcast channels but depends on alternating iteration procedure among several blocks of variables. Such an iterative nature suffers from high computational complexity, blocking the real-time implementation of traditional optimization-based beamforming solutions in practice.

Recently, deep learning (DL) techniques have been exploited for developing efficient beamforming strategies \cite{Xia-et-al:TC19, Huang-et-al:TVT20}. A key enabler of the DL-based beamforming schemes is to employ deep neural networks (DNNs), e.g., fully-connected neural networks (FNNs) and convolutional neural networks (CNNs), for shifting online calculations of traditional algorithms into their offline training processes with the aids of numerous channel state information (CSI) samples. It has been reported in various studies \cite{Samuel-et-al:TSP19}-\cite{Lee-et-al:JSAC19} that the DL-based wireless communication systems can achieve comparable (or even higher) performance to existing optimization approaches with much reduced online execution time. Similar results have recently been presented in \cite{Xia-et-al:TC19} and \cite{Huang-et-al:TVT20} where CNNs are utilized for optimizing beamforming vectors in multi-antenna downlink networks. The CNNs in \cite{Xia-et-al:TC19} and \cite{Huang-et-al:TVT20} accept CSI as an input feature and are trained to output efficient beamforming for a certain power constraint at base stations (BSs). The uplink-downlink duality has been revisited to develop efficient DNN architecture for improving various performance metrics such as the sum rate and the max-min signal-to-interference noise ratio (SINR).

The existing DL-based beamforming schemes have implicitly assumed that the transmit power budget levels at BSs are fixed. However, in practice, the power limitation should be regarded as stochastic number since it can vary according to propagation environment and network typologies. For instance, the transmit power budget is, in general, required to be set differently for individual frequency bands, since the carrier frequency affects the propagation properties of wireless signals \cite{Elsherif-et-al:JSAC15}. Such a scenario prevails in the carrier aggregation techniques in LTE-A systems. Also, the macro-cell BS is allowed to use a high transmit power, whereas a strict power constraint is typically incurred in a small-cell setup. Nevertheless, conventional beamforming solutions including the DL approaches \cite{Xia-et-al:TC19,Huang-et-al:TVT20} have focused on a fixed transmit power constraint and lack the adaptability for heterogeneous networking scenarios with arbitrary power budget.

This letter investigates a DL-based beamforming optimization strategy for the downlink of multi-user multi-antenna systems. We consider a heterogeneous network topology where the sum power constraint at a BS arbitrarily varies for the transmission environment. It is necessary to take a careful approach to the DL application to universally support various transmit power budget levels. New DL approaches not attempted in existing studies are presented. A key idea is to exploit the total transmit power budget as an additional input feature for a DNN so that it can effectively learn the impact of transmit power constraint in beamforming optimization. Such an approach enables us to focus on training a single DNN for a wide range of the power constraints, whereas the existing works \cite{Xia-et-al:TC19,Huang-et-al:TVT20} require multiple training tasks of DNNs designed for a certain choice of the power budget. This also leads to a memory-efficient implementation since the BS only requires to store a parameter set of a single DNN trained for all possible transmit power constraints. Since the performance of the DL techniques would depend on the DNN architecture, we investigate three different types of DNN structures suitable for the universal beamforming optimization. Our main focus is on developing the universal beamforming learning solution and comparing various DNN candidates with intensively examining the viability of the proposed DL methods not only in terms of the sum rate performance but also in terms of the computational complexity. Comparison among various DNN constructions for beamforming optimization has not yet been adequately studied in the literature. To this end, we present three different learning strategies including the concept of beam feature optimizations in \cite{Xia-et-al:TC19} as well as the direct learning of beam weights. Numerical results validate the advantages of the proposed universal DL approaches over existing methods only applicable to a fixed transmit power constraint. It is revealed that the FNN achieves almost identical performance of the CNN with reduced computational complexity.

% \vspace{-3mm}
\section{Multi-User MISO Downlink Systems\label{sec:System-Model}}

We consider MISO downlink systems where a BS with $M$ antennas serves $K$ single-antenna user equipments (UEs). Let $\mathcal{K}=\{1,2,\ldots,K\}$ be the set of UE indices. Denoting $\mathbf{h}_{k}\in\mathbb{C}^{M}$ as a channel vector from the BS to UE $k$, the SINR of UE $k$ is expressed by
% \vspace{-1mm}
\begin{equation}
\mathrm{SINR}_{k}(\mathbf{v})=\frac {|\mathbf{h}_{k}^{H}\mathbf{v}_{k}|^{2}}{\sum\nolimits _{j\in\mathcal{K}\setminus\{k\}}|\mathbf{h}_{k}^{H}\mathbf{v}_{j}|^{2}+1},\label{eq:Downlink-SINR}
% \vspace{-1mm}
\end{equation}
where $(\cdot)^H$ accounts for the Hermitian transpose, $\mathbf{v}_{k}\in\mathbb{C}^{M}$ stands for the beamforming vector for UE $k$, and $\mathbf{v}\triangleq[\mathbf{v}_{1}^{T}\cdots\mathbf{v}_{K}^{T}]^{T}\in\mathbb{C}^{MK}$ is a collection of the beamforming vectors with $(\cdot)^T$ equal to the transpose operation. We assume that the noise signals have unit power without the loss of generality. The BS is subject to a sum power constraint written as  $\sum\nolimits _{k\in\mathcal{K}}||\mathbf{v}_{k}||^{2}\leq P$ with $P$ being the maximum power budget. The level of the sum power constraint $P$ would rely on network typologies and can vary for each transmission. However, traditional algorithms can only be applied to a certain $P$, and thus we need to execute existing algorithms for all possible transmit power constraints.

The target of this work is to include various choices of $P$ into a single optimization formulation so that the resulting beamforming can be universally employed for heterogeneous networking scenarios with arbitrary $P$. Such a task can be formulated as the identification of a mapping $\mathbf{v}=\mathcal{V}(\mathbf{h},P)$ that characterizes an unknown computation strategy for the beamforming vectors based on the stacked CSI vector $\mathbf{h}\triangleq[\mathbf{h}_{1}^{T}\cdots\mathbf{h}_{K}^{T}]^{T}\in\mathbb{C}^{MK}$ as well as the sum power constraint $P$. Let $\mathcal{P}$ be the set of all possible transmit power levels. Then, the average sum rate maximization problem is given as
% \vspace{-1mm}
\begin{align}
\underset{\mathcal{V}(\cdot)}{\mathrm{max}}\,\, & \,\mathbb{E}_{\mathbf{h},P}\Bigg[\sum\nolimits _{k\in\mathcal{K}}\mathrm{log}_{2}\bigg(1+\mathrm{SINR}_{k}(\mathcal{V}(\mathbf{h},P))\bigg)\Bigg]\label{eq:problem-SR-maximization}\\
\mathrm{s.t.}\,\,\,\,\, & \sum\nolimits _{k\in\mathcal{K}}||\mathbf{J}_{k}\mathcal{V}(\mathbf{h},P)||^{2}= P,\ \forall\mathbf{h},\forall P\in\mathcal{P},\label{eq:constraint}%\\
% \vspace{-1mm}
\end{align}
where $\mathbb{E}_{X}[\cdot]$ accounts for the expectation operator over a random variable $X$ and $\mathbf{J}_{k}\in\mathbb{R}^{MK\times MK}$ stands for an all zero matrix whose $(M(k-1)+1)$-th to $(Mk)$-th diagonal elements being replaced with ones. The sum rate objective is averaged over the channel distribution as well as the sum power constraints so that the beamforming computation strategy $\mathcal{V}(\cdot)$ can be applied to a wide range of $P$. The power constraint in \eqref{eq:constraint} should be satisfied for an instantaneous choice of $P\in\mathcal{P}$. It is not straightforward to handle the non-convex formulation \eqref{eq:problem-SR-maximization} since no closed-form expressions are available both for the objective function and the mapping $\mathcal{V}(\cdot)$.

To overcome this challenge, we propose a DL method for addressing intractable formulation \eqref{eq:problem-SR-maximization}. There have been recent works on DL-based beamforming design \cite{Xia-et-al:TC19,Huang-et-al:TVT20}. However, they are developed for a specific $P$. This is crucial for a DL setup, since we need to train multiple DNNs for all possible candidates of $P$, which is obviously not practical due to the high computation power required for training processes. It is still unaddressed how to design a DNN structure and its learning strategy suitable for problem \eqref{eq:problem-SR-maximization}.

% \vspace{-2mm}
\section{Proposed DL Methods\label{sec:Deeplearning-approach}}

\begin{figure}
\centering\includegraphics[width=8.4cm,height=5.5cm]{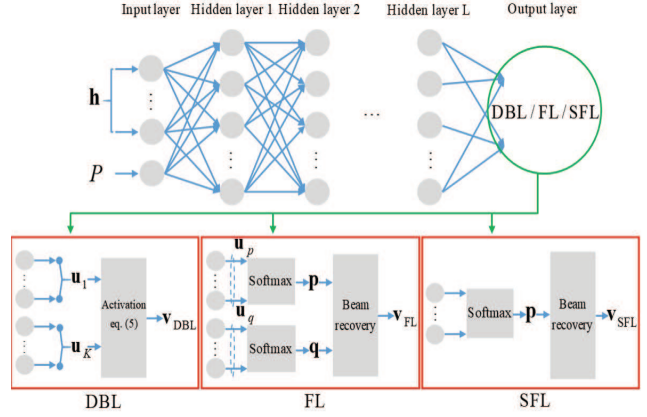}
% \vspace{-2.5mm}
\caption{{\footnotesize{}\label{fig:DNN_structure}Three types of DNN structure with $L$ hidden layers.}}
% \vspace{-4mm}
\vspace{-4mm}
\end{figure}

This section presents a DL-based beamforming optimization scheme that can be universally applied under arbitrary given sum power budget $P$. To this end, the mapping $\mathcal{V}(\mathbf{h},P)$ is replaced with a DNN illustrated in Fig. \ref{fig:DNN_structure} consisting of an input layer, $L$ hidden layers, and an output layer. For simplicity, the hidden layers are denoted by layer $l$ for $l=1,\cdots,L$, whereas the output layer is referred to as layer $L+1$. An input feature of the proposed DNN architecture is denoted by $\mathbf{x}_{0}$ and it becomes a concatenation of the CSI $\mathbf{h}$ and the power constraint as $\mathbf{x}_{0}\triangleq[\mathbf{h}^{T}, P]^{T}\in\mathbb{C}^{MK+1}$.\footnote{Since the current DL libraries such as Tensorflow do not support complex-valued calculation, real representations are employed for the implementation.} The proposed DNN can be an arbitrary feed-forward structure such as fully-connected and convolutional layers. Without loss of the generality, this section focuses on a FNN constructed only with fully-connected layers as illustrated in Fig. \ref{fig:DNN_structure}.\footnote{In Sec. \ref{sec:numerical_results}, we provide numerical results both for the FNN and CNN architectures.} A computation result of layer $l$ for $l=1,\cdots,L,L+1$ is denoted by $\mathbf{x}_{l}\in\mathbb{C}^{N_{l}}$ where $N_{l}$ accounts for the dimension of layer $l$. Then, the operation of layer $l$ is written by
% \vspace{-1.5mm}
\begin{equation}
\mathbf{x}_{l}= f_{l}\left(\mathbf{W}_{l}\mathbf{x}_{l-1}+\mathbf{b}_{l}\right), \label{eq:x_l}
% \vspace{-1.5mm}
\end{equation}
where an element-wise mapping $f_{l}(\cdot)$ stands for an activation function and $\mathbf{W}_{l}\in\mathbb{C}^{N_{l}\times N_{l-1}}$ and $\mathbf{b}_{l}\in\mathbb{C}^{N_{l}}$ respectively indicate an weight matrix and a bias vector at layer $l$. The final output $\mathbf{x}_{L+1}$ is then exploited as a stacked beamforming vector $\mathbf{v}$. Let $\Theta = \{\mathbf{W}_{l}, \mathbf{b}_{l}\}_{l=1,\ldots,L+1}$ be the DNN parameter set collecting the weight matrices and bias vectors. Then, the overall forward pass procedure of the DNN is parmeterized as $\mathbf{v}=\mathcal{F}(\mathbf{x}_{0};\Theta)$, where a vector function $\mathcal{F}(\cdot;\Theta)$ characterizes the input-output relationship of the DNN specified by consecutive neural computations \eqref{eq:x_l}.

We apply the DNN $\mathcal{F}(\cdot;\Theta)$ to approximate the unknown beamforming calculation strategy $\mathcal{V}(\cdot)$. The accuracy of the DNN approximation has been verified by the universal approximation theorem \cite{Hornik-et-al:NN89}. However, it only guarantees the existence of the DNN with a certain level of the approximation accuracy, but not provides any guidelines for the DNN construction. Therefore, it is essential to design an efficient DNN which can successfully model the optimal beamforming computation rule $\mathcal{V}(\cdot)$ for \eqref{eq:problem-SR-maximization}. To this end, we need to determine features of the layers, i.e., the activation $f_{l}(\cdot)$, dimension $N_{l}$, and the number of the hidden layers $L$. In the case of the hidden layers, these hyper-parameters are, in general, selected via trial-and-error processes based on the validation performance
\cite{Lee-et-al:JSAC19}. In contrast, the output layer should be carefully designed since it is directly related to the solution of the constrained formulation \eqref{eq:problem-SR-maximization}. In the following, we investigate various candidates for the output layer, and propose three different DL approaches named {\em Direct Beamforming Learning (DBL)}, {\em Feature Learning (FL)}, and {\em Simplified Feature Learning (SFL)}.

% \vspace{-3mm}
\subsection{Direct Beamforming Learning\label{subsec:Directly-learning}}

In the DBL scheme, the DNN outputs a beamforming vector directly. Thus, the output activation should be carefully designed to always yield a feasible solution satisfying the sum power constraint \eqref{eq:constraint}. Let $\mathbf{u}\triangleq\mathbf{W}_{L+1}\mathbf{x}_{L}+b_{L+1}\in\mathbb{C}^{MK}$ with $\mathbf{u}=[\mathbf{u}_{1}^{T}\cdots\mathbf{u}_{K}^{T}]^{T}$ be a vector to be fed to the activation of the output layer. Then, the beamforming vector $\mathbf{v}_{\text{DBL},k}$ of the DBL method is determined based on $\mathbf{u}_{k}$ as
% \vspace{-1mm}
\begin{equation}
\mathbf{v}_{\text{DBL},k}=\sqrt{\frac {P} {\sum\nolimits _{j\in\mathcal{K}}||\mathbf{u}_{j}||^{2}}} \mathbf{u}_{k}, \label{eq:tilde_v_k}
% \vspace{-1mm}
\end{equation}
where the output activation of the DBL can be specified by \eqref{eq:tilde_v_k}. It is not difficult to see that $\mathbf{v}_{\text{DBL},k}$ fulfils the sum power constraint $\sum_{k\in\mathcal{K}}\|\mathbf{v}_{\text{DBL},k}\|^{2}=P$. Consequently, the DNN mapping of the DBL is denoted by $\mathbf{v}_{\text{DBL}}=\mathcal{F}_{\text{DBL}}(\mathbf{x}_{0};\Theta_{\text{DBL}})$ with $\mathbf{v}_{\text{DBL}}\triangleq[\mathbf{v}_{\text{DBL},1}^{T}\cdots\mathbf{v}_{\text{DBL},K}^{T}]^{T}$ and a parameter set $\Theta_{\text{DBL}}$.

% \vspace{-3mm}
\subsection{Feature Learning\label{subsec:Learning-Xia}}
The DBL approach may encounter the overfitting problem with a poor performance since the number of variables to be predicted increases both with the number of BS antennas $M$ and UEs $K$. To handle this issue, we study an alternative strategy named FL which produces low-dimensional intermediate variables that act as key feature for constructing the optimal beamforming solution. The beamforming vector $\mathbf{v}_{\text{FL},k}$ of the FL method is determined based on the optimal beamforming structure resulting from the uplink-downlink duality as  \cite{Boche-et-al:SA05}
% \vspace{-1mm}
\begin{equation}
\mathbf{v}_{\text{FL},k}\!=\!\sqrt{p_{k}} \frac {(\mathbf{I}_{M}+\sum_{j\in\mathcal{K}}q_{j}\mathbf{h}_{j}\mathbf{h}_{j}^{H})^{-1}\mathbf{h}_{k}} {||(\mathbf{I}_{M}+\sum_{j\in\mathcal{K}}q_{j}\mathbf{h}_{j}\mathbf{h}_{j}^{H})^{-1}\mathbf{h}_{k}||}
\!\triangleq\!\sqrt{p_{k}}\mathbf{d}_{k}, \label{eq:Optimal-BF-structure}
% \vspace{-1mm}
\end{equation}
where $\mathbf{I}_{m}$ accounts for the $m$-by-$m$ identity matrix; and $p_{k}$ and $q_{k}$ reflect the primal downlink power and the dual uplink power for UE $k$, respectively, which are subject to the sum power constraints $\sum\nolimits _{k\in\mathcal{K}}p_{k}=\sum\nolimits _{k\in\mathcal{K}}q_{k}=P$; and $\mathbf{d}_{k}$ is the normalized beamforming vector, that determines the direction of $\mathbf{v}_{\text{FL},k}$. In \eqref{eq:Optimal-BF-structure}, the optimization variables now boil down into $2K$ real numbers $p_{k}$ and $q_{k}$, $\forall k\in\mathcal{K}$, which is much smaller than identifying $2MK$ beam weights in the DBL. The scale variables $p_{k}$ and $q_{k}$ which are directly passed to the beam recovery activation can be obtained through the action of the output layer. The FL has been recently introduced in \cite{Xia-et-al:TC19} for finding downlink MISO beamforming vectors. However, the advantages of the FL structure in terms of the learning performance and running time have not been adequately investigated in comparison with other DNN architectures including the DBL strategy. Thereby, it is still unclear that the FL indeed performs better than the DBL that learns the beamforming straightforwardly. Furthermore, the existing work has focused on the case where the sum power constraint $P$ is fixed. Therefore, a modified design for the output activation is essential for the universal beamforming scheme.

We design the output layer of the DNN suitable for the FL when $P$ varies for each training sample. With a slight abuse of notations, an input to the output activation is denoted by $\mathbf{u}=[\mathbf{u}_{p}^{T},\mathbf{u}_{q}^{T}]^{T}\in\mathbb{R}^{2K}$ where $\mathbf{u}_{p}\in\mathbb{R}^{K}$ and $\mathbf{u}_{q}\in\mathbb{R}^{K}$ will be utilized to extract the downlink power $\mathbf{p}\triangleq[p_{1}\cdots p_{K}]^{T}$ and the dual uplink power $\mathbf{q}\triangleq[q_{1}\cdots q_{K}]^{T}$, respectively. We first apply a scaled softmax function $\text{softmax}_{P}(\mathbf{z})_{i}=\frac {Pe^{z_{i}}} {\sum\nolimits_{k\in\mathcal{K}}e^{z_{k}}}, \forall i\in\mathcal{K}$ and $\mathbf{z}\triangleq[z_{1}, \ldots, z_{K}]\in\mathbb{R}^{K}$ to individual $\mathbf{u}_{p}$ and $\mathbf{u}_{q}$ for obtaining feasible $\mathbf{p}$ and $\mathbf{q}$ such that $\sum\nolimits _{k\in\mathcal{K}}p_{k}=\sum\nolimits _{k\in\mathcal{K}}q_{k}=P$ (See Fig. \ref{fig:DNN_structure}). Then, we retrieve the beamforming vector $\mathbf{v}_{\text{FL},k}$ of the FL method from the optimal beam structure in \eqref{eq:Optimal-BF-structure}. Finally, the DNN mapping of the FL approach can be expressed by $\mathbf{v}_{\text{FL}}=\mathcal{F}_{\text{FL}}(\mathbf{x}_{0};\Theta_{\text{FL}})$ with a parameter set $\Theta_{\text{FL}}$ and the resulting beamforming $\mathbf{v}_{\text{FL}}=[\mathbf{v}_{\text{FL},1}^{T}\cdots\mathbf{v}_{\text{FL},K}^{T}]^{T}$ whose output activation becomes an integration of the softmax and the beam recovery operation~\eqref{eq:Optimal-BF-structure}. The FL would require more computation in the output activation than the DBL approach due to the matrix inversion~in~\eqref{eq:Optimal-BF-structure}.

% \vspace{-3mm}
\subsection{Simplified Feature Learning\label{subsec:Proposed-learning}}
We propose a simplified method that further reduces the output dimension of the FL. A key idea is to construct a DNN which can recovers the dual uplink power variable $\mathbf{q}$ from the downlink power variable $\mathbf{p}$. In \cite{Boche-et-al:SA05}, it has been revealed that the power vectors achieving a certain SINR target $\text{SINR}_{k}(\mathbf{v})\geq\gamma_{k}$, $\forall k\in\mathcal{K}$, are expressed as \eqref{eq:Optimal-BF-structure} with $\mathbf{p}=\sigma^{2}\mathbf{\Omega}^{-1}\mathbf{1}_{K}$ and $\mathbf{q}=\sigma^{2}(\mathbf{\Omega}^{T})^{-1}\mathbf{1}_{K}$, where $\mathbf{1}_{m}$ stands for an all-one vector of length $m$ and the $(k,j)$-th element $[\mathbf{\Omega}]_{kj}$ of the matrix $\mathbf{\Omega}\in\mathbb{R}^{K\times K}$ is given by
% \vspace{-1mm}
\begin{align}
[\mathbf{\Omega}]_{kj}=\begin{cases}-\frac{1}{\gamma_{k}}|\mathbf{h}_{k}^{H}\mathbf{d}_{k}|^{2}, & k = j,\\ |\mathbf{h}_{k}^{H}\mathbf{d}_{j}|^{2}, & k \neq j,\end{cases} \label{eq:Omega-matrix}
% \vspace{-3mm}
\end{align}
with $\mathbf{d}_{k}$ defined in \eqref{eq:Optimal-BF-structure}. The off-diagonal element $|\mathbf{h}_{k}^{H}\mathbf{d}_{j}|^{2}$ of the matrix $\mathbf{\Omega}$ corresponds to the normalized inter-user interference at UE $k$ stemmed from UE $j$. With the optimal beam structure \eqref{eq:Optimal-BF-structure}, the interference power becomes negligible compared to the desired signal power, i.e., the diagonal element of $\mathbf{\Omega}$. In particular, at the high signal-to-noise (SNR) regime, the interference power goes to zero since the zero-forcing (ZF) transmission becomes a near-optimal solution for the sum rate maximization \cite{Bjornson-et-al:SPM14}. This motivates us to employ an approximation $\mathbf{\Omega}\simeq\mathbf{\Omega}^{T}$, leading to $\mathbf{p}=\sigma^{2}\mathbf{\Omega}^{-1}\mathbf{1}_{K}\simeq\sigma^{2}(\mathbf{\Omega}^{T})^{-1}\mathbf{1}_{K}=\mathbf{q}$.

Based on this observation, we provide a SFL approach which only learns the downlink power vector $\mathbf{p}$. Compared to the FL that finds $2K$ variables for both the uplink and downlink powers, the output dimension of the SFL becomes only $K$, which would lead to a more efficient training procedure. The scaled softmax function $\text{softmax}_{P}(\mathbf{z})$ is applied to the output layer to yield a feasible $\mathbf{p}$ satisfying $\sum_{k\in\mathcal{K}}p_{k}=P$. Then, the beamforming vector $\mathbf{v}_{\text{SFL},k}$ of the SFL method is obtained by substituting $\mathbf{q}=\mathbf{p}$ into \eqref{eq:Optimal-BF-structure}. Similar to the FL, the output activation of the SFL strategy is determined by the scaled softmax followed by the beam computation \eqref{eq:Optimal-BF-structure}. The resulting DNN for the SFL is denoted by $\mathbf{v}_{\text{SFL}}=\mathcal{F}_{\text{SFL}}(\mathbf{x}_{0};\Theta_{\text{SFL}})$ with $\mathbf{v}_{\text{SFL}}\triangleq[\mathbf{v}_{\text{SFL},1}^{T}\cdots\mathbf{v}_{\text{SFL},K}^{T}]^{T}$.

% \vspace{-3mm}
\subsection{Training and Implementation}
We discuss the training process of the proposed DL methods. It is worth noting that all the learning techniques presented in the previous sections are carefully designed to meet the sum power constraint \eqref{eq:constraint}. Applying $\mathcal{V}(\mathbf{h},P)=\mathcal{F}_{S}(\mathbf{x}_{0};\Theta_{S})$ to \eqref{eq:problem-SR-maximization} for each scheme $S\in\{\text{DBL},\text{FL},\text{SFL}\}$, an unconstrained training problem can be formulated as
% \vspace{-1mm}
\begin{align}
\underset{\Theta_{S}}{\mathrm{max}}\,\, & \,\mathbb{E}_{\mathbf{h}, P}\Bigg[\sum\nolimits _{k\in\mathcal{K}}\mathrm{log}_{2}\bigg(1+\mathrm{SINR}_{k}(\mathcal{F}_{S}(\mathbf{x}_{0};\Theta_{S}))\bigg)\Bigg].\label{eq:train}
% \vspace{-3mm}
\end{align}
The optimization variable is now transformed into the DNN parameter $\Theta$. The solution of \eqref{eq:train} can be obtained by the state-of-the-art mini-batch stochastic gradient decent (SGD) algorithms, e,g., the Adam algorithm \cite{Kingma-et-al:ICLR15}. At the $n$-th iteration of the SGD algorithm, the DNN parameter is updated as
% \vspace{-1mm}
\begin{align} \label{eq:mini-batch-SGD}
\Theta_{S}^{[n]}&=\Theta_{S}^{[n-1]}\\&+\eta\mathbb{E}_{\mathcal{B}}\Bigg[  \sum_{k\in\mathcal{K}}\nabla_{\Theta_{S}}\mathrm{log}_{2}\Big(1+\mathrm{SINR}_{k}\big(\mathcal{F}_{S}(\mathbf{x}_{0};\Theta_{S}^{[n-1]})\big)\Big)\Bigg],  \nonumber
% \vspace{-3mm}
\end{align}
where $\nabla_{\mathbf{z}}f(\mathbf{z})$ indicates the gradient of a function $f(\mathbf{z})$ with respect to $\mathbf{z}$, $\Theta_{S}^{[n]}$ stands for the DNN parameter at the $n$-th iteration, $\eta>0$ is the learning rate, $\mathcal{B}$ denotes the mini-batch set containing $B$ independent samples of the DNN input $(\mathbf{h},P)$. In practice, the state-of-art DL libraries, e.g., Tensorflow and Pytorch, numerically calculate the gradient in \eqref{eq:mini-batch-SGD} through the backpropagation algorithm \cite{Goodfellow-et-al} based on the chain rule. The training strategy in \eqref{eq:mini-batch-SGD} does not require labels for the input feature $[\mathbf{h}^{T},P]^{T}$, i.e., the prior knowledge regarding the beamforming solution. Therefore, the proposed DL approaches can be viewed as the unsupervised learning framework where the DNNs are trained to directly maximize the average sum rate performance, but not to memorize existing solution as in the supervised learning method \cite{Sun-et-al:TSP18}. Note that the training is performed in the offline domain by collecting numerous training samples in advance. The real-time inference of the trained DNN is simply carried out by the linear matrix multiplications \eqref{eq:x_l} with the trained DNN parameter stored into memory units of the BS.

Next, we discuss the computational complexity of three different DL designs, focusing on the testing process performed in the online domain. The forward pass computations at the hidden layers of the DBL, FL, and SFL are the same, and the difference comes from their output layer computations, i.e., \eqref{eq:tilde_v_k} and \eqref{eq:Optimal-BF-structure}. The output layer of the DBL method retrieves the beamforming vector by using \eqref{eq:tilde_v_k}, requiring the complexity $\mathcal{O}(KM)$. In the FL and the SFL, the complexity of the beam recovery process in \eqref{eq:Optimal-BF-structure} is given by $\mathcal{O}(KM^{2}+M^{3})$ \cite{Xia-et-al:TC19}, which is more complicated than the DBL due to the matrix inversion.

%\vspace{-1mm}
\section{Numerical Results}\label{sec:numerical_results}
This section assesses the performance of the proposed DL methods. The UEs are randomly dropped within a cell of radius $100\ \mathrm{m}$ and the BS is fixed at the center of the cell. The channel is modeled as $\mathbf{h}_{k}=\sqrt{\rho_{k}}\mathbf{\tilde{h}}_k$ where
$\rho_{k}=\frac{1}{1+(d_{k}/d_{\mathrm{0}})^{\alpha}}$ and
$\tilde{\mathbf{h}}_{k}\sim\mathcal{CN}(\mathbf{0},\mathbf{I}_{M})$ stand for long-term path-loss and small-scale fading, respectively. Here $d_{k}$ is the distance between the BS and UE $k$, $d_{\mathrm{0}}=30\ \mathrm{m}$ equals the reference distance, and $\alpha=3$ indicates the path-loss exponent. Since the additive noise variance is set to unity, the SNR is defined as $P$. We consider both FNN and CNN structures for the DNN construction. The FNN has $L=5$ fully-connected hidden layers with a rectified linear unit (ReLU) activation $f_{l}(z)=\max\{0,z\}$ and output dimension $320$. On the other hand, the CNN is designed as an integration of three convolutional layers with $60$ kernels of size $3\times3$ and subsequent two fully-connected layers with the output dimension $320$. The Adam algorithm is adopted with the learning rate $\eta = 10^{-3}$ and the mini-batch size $10^{4}$. Each mini-batch contains independently generated CSI as well as the unbalanced samples from the uniformly distributed transmit power $P\in\{0\ \text{dB},5\ \text{dB},\cdots,30\ \text{dB}\}$. The batch normalization \cite{loffe-et-al:ICML15} follows each layer to accelerate training processes. The sum rate performance is evaluated over $10^{3}$ test samples. The training is implemented with Tensorflow on a PC equipped with an Intel i7-7700K CPU, 64 GB of RAM, a Titan XP GPU.

\begin{table}
\centering\caption{\label{tab:table1}Average Sum Rate Performance [bps/Hz].}
\begin{tabular}{|c|c|c|c|c|c|c|}
\hline
\multirow{2}{*}{} & \multicolumn{2}{c|}{DBL} & \multicolumn{2}{c|}{FL} & \multicolumn{2}{c|}{SFL} \\ \cline{2-7}
 & FNN & CNN & FNN & CNN & FNN & CNN \\ \hline \hline
$P=0$ dB & 1.20 & 1.17 & 1.23 & 1.23 & 1.23 & 1.23 \\ \hline
$P=10$ dB & 3.95 & 3.81 & 4.13 & 4.15 & 4.14 & 4.14 \\ \hline
$P=20$ dB & 8.37 & 8.21 & 9.83 & 9.86 & 9.82 & 9.85 \\ \hline
$P=30$ dB & 12.47 & 12.84 & 18.37 & 18.35 & 18.37 & 18.43 \\ \hline
\end{tabular}
\vspace{-4.7mm}
\end{table}

\begin{table*}[htp!]
\centering\caption{\label{tab:table2}Average CPU Running Time [sec].}
\subcaption {$M=K=4$}
\begin{tabular}{|c|c|c|c|c|c|c|c|c|}
\hline
\multicolumn{2}{|c|}{DBL} & \multicolumn{2}{c|}{FL} & \multicolumn{2}{c|}{SFL} & \multicolumn{3}{c|}{WMMSE} \\ \hline \hline
FNN & CNN & FNN & CNN & FNN & CNN & $P=0$ dB & $P=10$ dB & $P=20$ dB \\ \hline
1.952e-4 & 1.161e-3 & 4.139e-4 & 1.393e-3 & 4.074e-4 & 1.385e-3 & 4.519e-3 & 9.767e-3 & 3.476e-2 \\ \hline
\end{tabular}
\vspace{2.5mm}
\subcaption{$M=K=6$}
\begin{tabular}{|c|c|c|c|c|c|c|c|c|}
\hline
\multicolumn{2}{|c|}{DBL} & \multicolumn{2}{c|}{FL} & \multicolumn{2}{c|}{SFL} & \multicolumn{3}{c|}{WMMSE} \\ \hline \hline
FNN & CNN & FNN & CNN & FNN & CNN & $P=0$ dB & $P=10$ dB & $P=20$ dB \\ \hline
2.661e-4 & 2.601e-3 & 5.525e-4 & 2.816e-3 & 5.461e-4 & 2.806e-3 & 8.537e-3 & 1.991e-2 & 7.002e-2 \\ \hline
\end{tabular}
\vspace{-5mm}
\end{table*}

\begin{figure}
$\!\!\!$%
\begin{minipage}[t]{0.45\columnwidth}%
\centering\includegraphics[width=4.8cm,height=5.2cm]{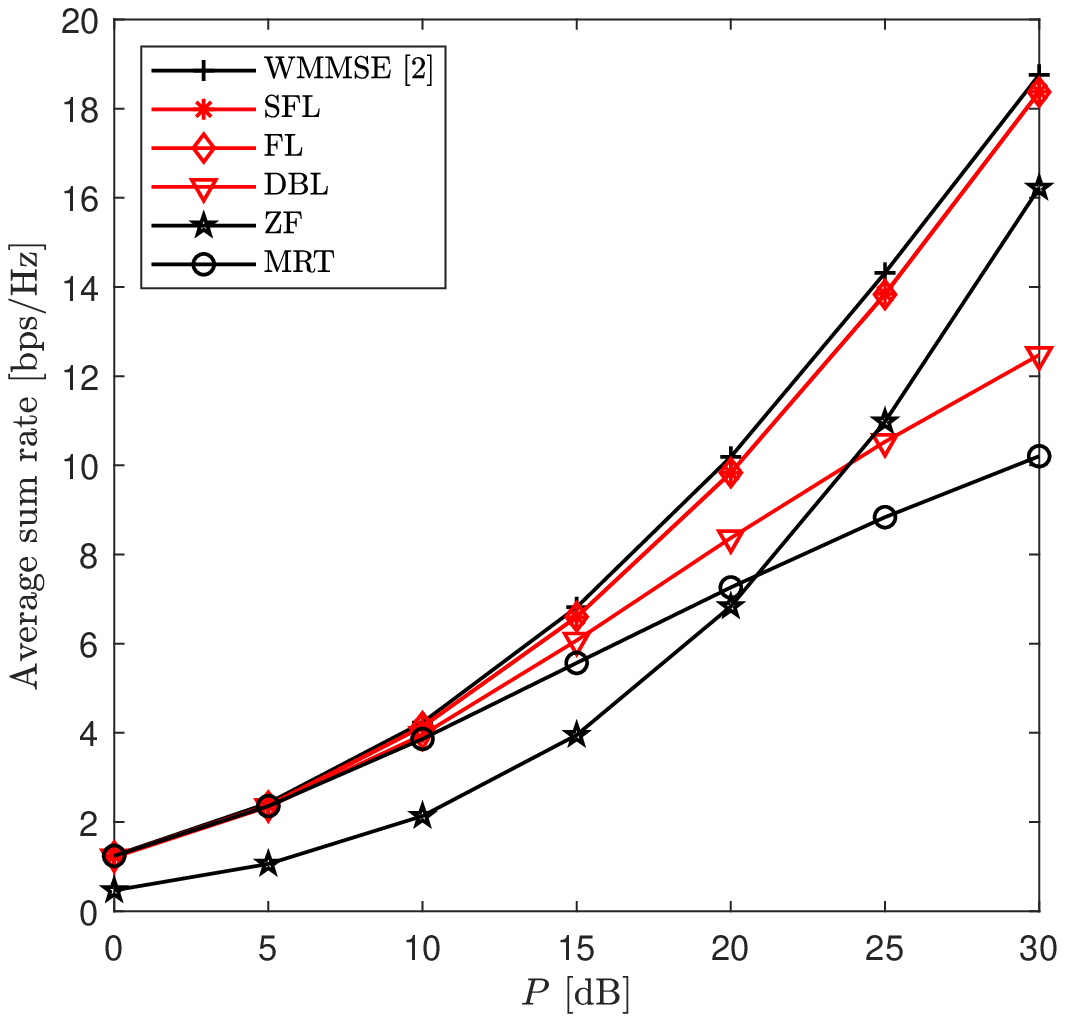}

\centering~~~~~~~{\footnotesize{}(a) $M=K=4$}{\footnotesize \par}%
\end{minipage}~~~~%
$\quad\!\!\!\!\!\!$
\begin{minipage}[t]{0.45\columnwidth}%
\centering\includegraphics[width=4.8cm,height=5.2cm]{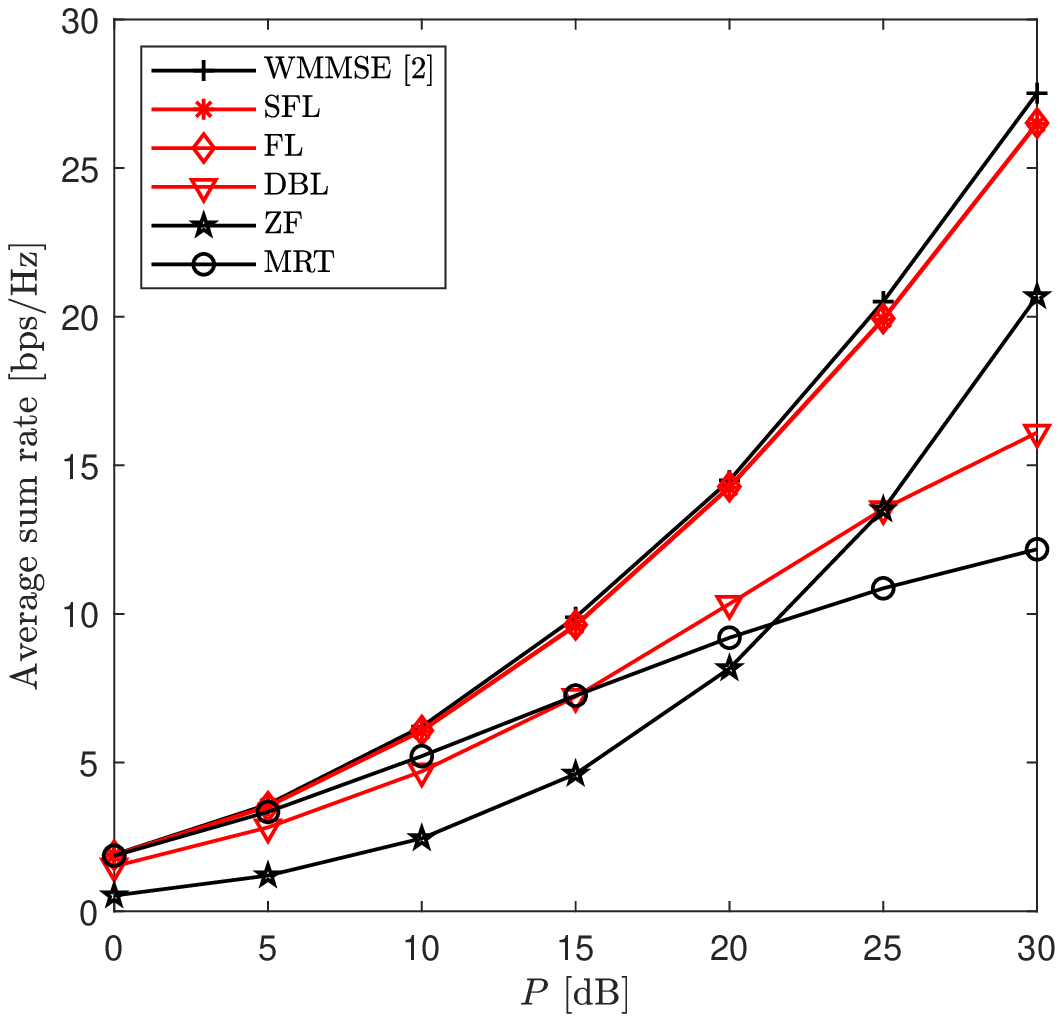}

\centering~~~~~~~{\footnotesize{}(b) $M=K=6$}{\footnotesize \par}%
\end{minipage}

\caption{\label{fig:graph1}Average sum rate with respect to $P$.}
\vspace{-6.5mm}
\end{figure}

We first focus on the FNN structure to verify the effectiveness of the proposed universal DL approach. Fig. \ref{fig:graph1} depicts the average sum rate performance of the proposed DL schemes with $M=K=4$ and $M=K=6$. As a benchmark, a locally optimum performance is plotted by employing the WMMSE algorithm \cite{Christensen-et-al:TWC08}. We also exhibit the performance of the MRT and the ZF transmission strategies with the optimized transmit power. Unlike the benchmark schemes that should be executed for each $P$, a more efficient implementation is possible for the proposed DL methods since they can be applied universally to all simulated range of $P$. Nevertheless, the proposed FL and SFL approaches achieve the almost identical performance to the WMMSE solution. In contrast, the performance of the DBL method degrades especially in the high SNR regime, proving the advantages of the data-driven learning strategy for the model-based DNN structure \eqref{eq:Optimal-BF-structure}. The gap between the FL and the SFL is negligible for all SNR range. This implies that the approximation $\mathbf{\Omega}\simeq\mathbf{\Omega}^{T}$ exploited for the SFL approach is effective for extracting useful features to construct the near-optimal beamforming solution.

\begin{figure}
$\!\!\!$%
\begin{minipage}[t]{0.45\columnwidth}%
\centering\includegraphics[width=4.8cm,height=5.2cm]{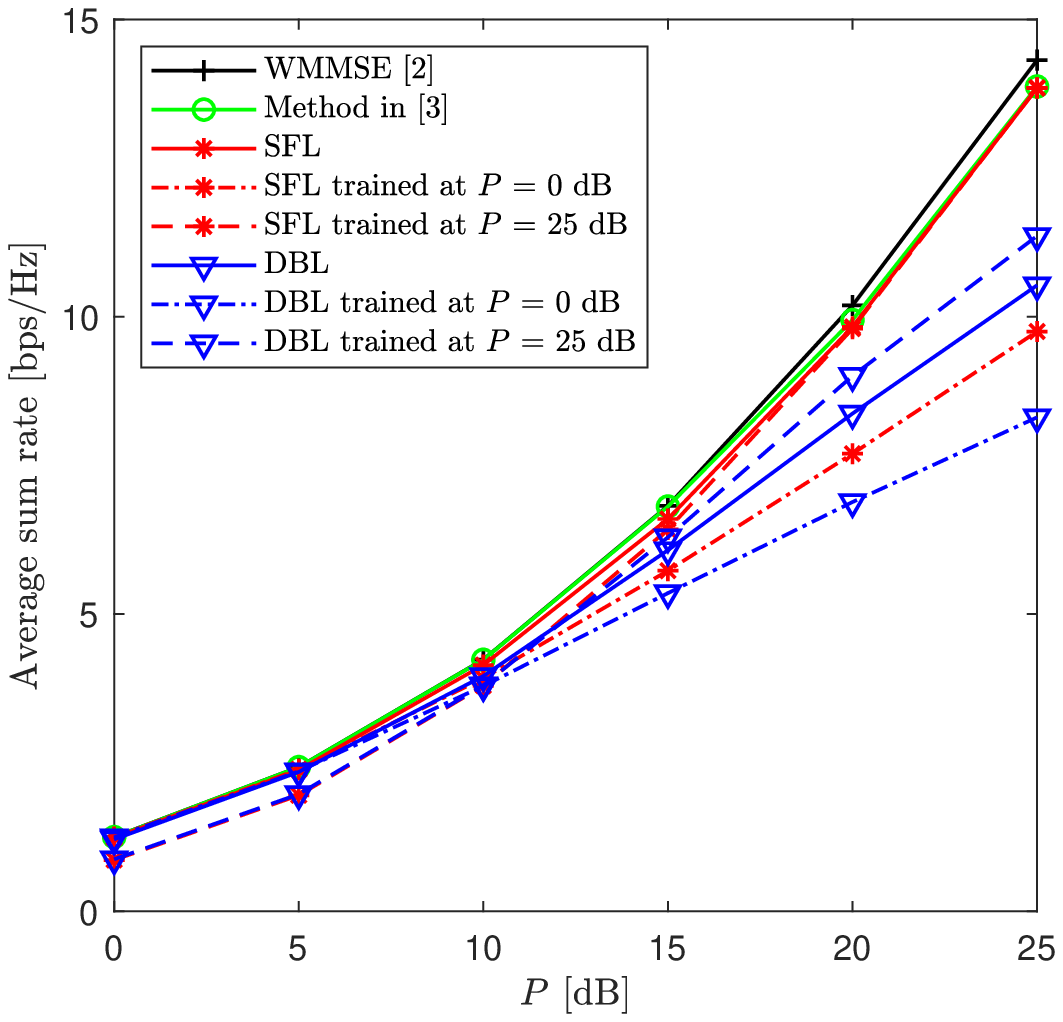}

\centering~~~~~~~{\footnotesize{}(a) $M=K=4$}{\footnotesize \par}%
\end{minipage}~~~~%
$\quad\!\!\!\!\!\!$
\begin{minipage}[t]{0.45\columnwidth}%
\centering\includegraphics[width=4.8cm,height=5.2cm]{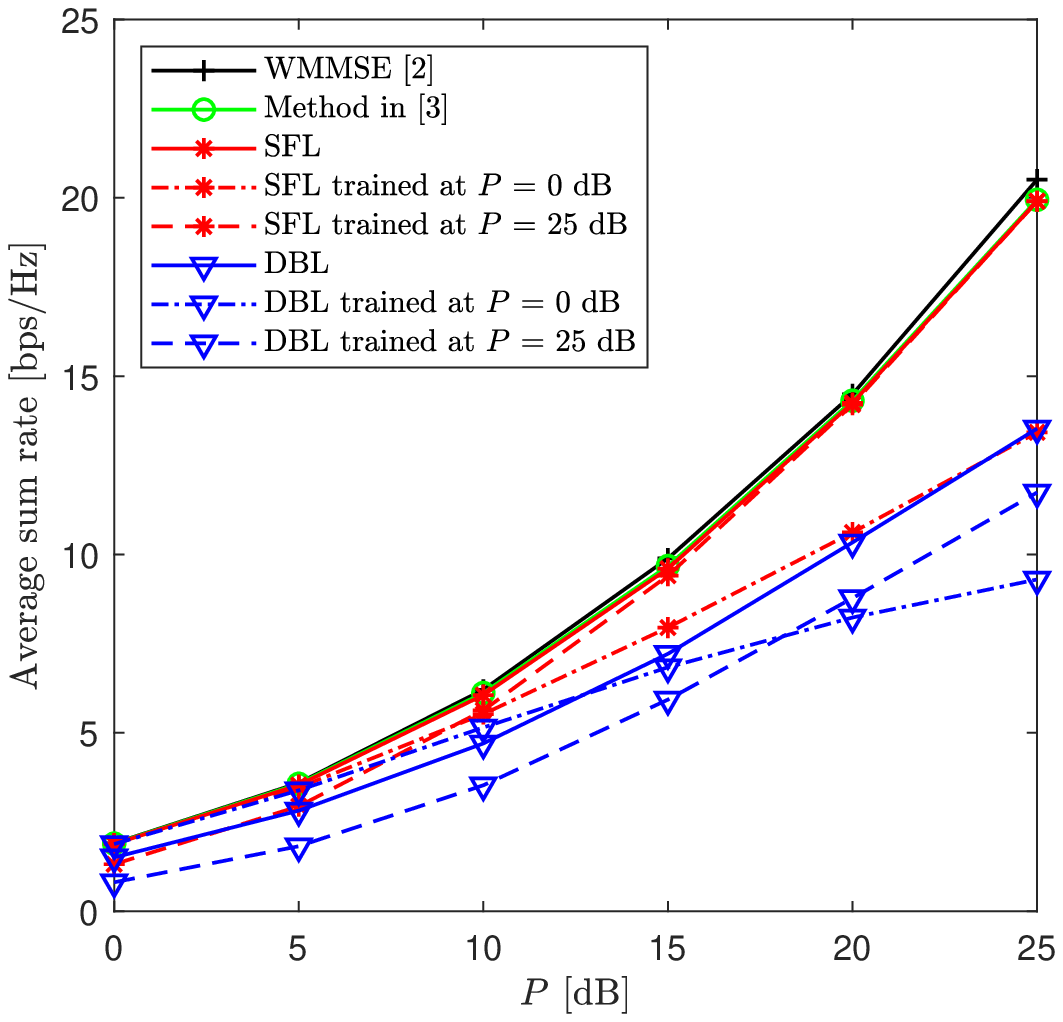}

\centering~~~~~~~{\footnotesize{}(b) $M=K=6$}{\footnotesize \par}%
\end{minipage}

\caption{\label{fig:graph2}Impact of the proposed universal DL approach.}
\vspace{-6.5mm}
\end{figure}

To see the impact of the universality of the proposed DL approach, Fig. \ref{fig:graph2} presents the average sum rate performance for $M = K = 4$ and $M = K = 6$ when the FNNs are trained at a specific $P$ ($P=0$ and $25$ dB) without inputting it to FNN but tested over a wide range of $P$. We also include the conventional method in \cite{Xia-et-al:TC19} where the stochastic variable $P$ is set to be equal in the training and testing of an FNN with the FL strategy. The universally trained SFL achieves almost identical performance to the non-universal FL \cite{Xia-et-al:TC19} that needs to train six different FNNs for each power level. This implies that the proposed universal DL schemes, which only requires a single training process, could reduce the training cost as well as the memory requirements without sacrificing the sum rate performance. Unlike the universally trained SFL, those trained at $P=0$ and $25$ dB exhibit degraded performance, showing the impact of $P$ on the training process. Thus, the universal DL approach is essential for practical systems where the power budget would be different for transmission scenarios.

Next, we examine the effect of the neural network structure on the performance. Table \ref{tab:table1} compares the average sum rate performance for $M=K=4$ when the proposed DL methods are implemented with the FNNs and the CNNs. It is seen that the CNN performs better than the FNN architecture, but the gap is negligible. This implies that the fully-connected structure is sufficient to obtain a good beamforming vector for maximizing the average sum rate performance.

Lastly, in Table \ref{tab:table2}, we present the average CPU execution time of various beamforming schemes for $M=K=4$ and $M=K=6$. Regardless of the configurations and the neural network structures, the proposed DL methods remarkably reduce the computation time of the conventional WMMSE algorithm. We can see that the time complexity of the CNN is higher than that of the FNN. Combining this with the results in Table \ref{tab:table1}, it can be concluded that the FNN would be an efficient DNN candidate achieving a good trade-off between the performance and the online computation time. As expected, both for the FNN and CNN, the DBL approach requires the lowest running time than other DL schemes. The WMMSE algorithm requires more time for convergence as $P$ gets larger, whereas the computation times of the proposed DL methods do not depend on $P$. Such a very important interpretation is observed equally regardless of the numbers of BS antennas and UEs.

%\vspace{-1mm}
\section{Conclusion\label{sec:Conclusion}}

This letter has studied a DL approach to optimize the beamforming vector that can be universally applicable for arbitrarily given power budget at the BS. To this end, the transmit power constraint at the BS has been exploited as side information so that the DNNs can efficiently learn the intractable mapping from the power budget to the optimal beamforming. Three different DNN structures have been presented. Numerical results have confirmed the advantages of the proposed universal DNN approach over conventional schemes.

%\vspace{-1mm}
%\begin{group}
\renewcommand{\baselinestretch}{0.90}

%\end{group}

\end{document}